\documentclass{JHEP3}
\usepackage{amsmath}
\usepackage[mathscr]{eucal}
\usepackage{epsfig}

\newcommand{\Q}{{\cal Q}}

\newcommand{\m} {\operatorname{M}}
\newcommand{\NM} {\operatorname{NM}}

\newcommand{\bos} {\operatorname{bos}}
\newcommand{\mat} {\operatorname{mat}}
\newcommand{\gh} {\operatorname{gh}}

\newcommand{\id} {{\cal I}}
\newcommand{\cA} {{\cal A}_W}
\newcommand{\cAA} {{\cal A}}

\newcommand{\cN} {{\cal N}}
\newcommand{\cn} {{\mathscr N}}

\newcommand{\cO} {{\cal O}}
\newcommand{\al} {{\alpha}}
\newcommand{\la} {{\lambda}}
\newcommand{\La} {{\Lambda}}
\newcommand{\ep} {{\epsilon}}

\newcommand{\ket}[1] {\left|#1\right>}

\newcommand{\vev}[1] {\left<#1\right>}

\title{On string fields and superstring field theories}
\author{Michael Kroyter\\
Center for Theoretical Physics\\
Massachusetts Institute of Technology\\
Cambridge, MA 02139, USA\\
\\ and\\ \\
School of Physics and Astronomy\\
The Raymond and Beverly Sackler Faculty of Exact Sciences\\
Tel Aviv University, Ramat Aviv, 69978, Israel\\ \\
\email{mikroyt@mit.edu}, \email{mikroyt@tau.ac.il}}

\abstract{We offer some thoughts regarding the space of string fields.
We suggest that this space should be identified as the odd component
of a star-algebra and focus among other issues on the role of the mid-point.
We argue that theories with mid-point insertions in the action, such as
the modified cubic theory can be well behaved, even if this mid-point
insertion has a non-trivial kernel.

We then discuss the recent proposal by Berkovits and Siegel of a non-minimal
superstring field theory. In this theory the action contains a mid-point
insertion of
a non-zero conformal weight. We show that, while this is a-priori a problem,
it might be possible (in the NS sector) to make sense out of this theory by
regularizing it. A cleaner resolution of the problem is to extend the
non-minimal sector in a way that allows a zero-weight mid-point insertion
with the desired properties.
We also study the generalisation of the theory to the
NS$-$ sector and explain the problems with defining the Ramond sectors.

We show that the non-minimal theory supports all the known solutions of the
standard modified cubic superstring field theory, including the
GSO+ vacuum solution. The properties of the solutions
carry over to the non-minimal theory. In particular, the vacuum solution has
the correct tension and cohomology.}

\preprint{MIT-CTP-4027\\TAUP-2894-09\\NSF-KITP-09-33}

\begin{document}

%=====================
\section{Introduction}
%=====================

The first attempt at constructing a covariant open superstring field theory
was made by Witten~\cite{Witten:1986qs}, following his construction of the
bosonic theory~\cite{Witten:1986cc}\footnote{For a detailed
introduction to string field theory and recent results within it one can
consult~\cite{Fuchs:2008cc}. Section 8 therein contains an introduction to
superstring field theories.}.
The most salient feature of the RNS superstring as compared to the bosonic
theory is the redundancy of vertex operators: Each state can be represented
by infinitely many vertex operators that differ by a new quantum number
called picture number~\cite{Friedan:1985ge}. CFT expectation values are
non-zero only for $-2$ picture number.
Witten suggested to work in the NS sector with string fields whose
picture number is ``the natural'' $-1$ picture. This implied that a picture
changing operator should be inserted on the interaction term. Associativity
of the star-product, gauge symmetry of the theory as well as the global
residual conformal symmetries generated by the $K_n$
operators~\cite{Witten:1986qs}, imply that the picture changing operator
has to be inserted at the string mid-point.

The string mid-point is invariant under the star product\footnote{Throughout
this work the star product is implicit. It should cause no confusion, as
there is no other way to multiply string fields. Another convention we are
using is writing $[\cdot,\cdot]$ for the graded-commutator.}. For
Witten's theory this implies that picture changing operators
collide when one considers gauge transformations or string
scattering~\cite{Wendt:1987zh}. These collisions render the theory
inconsistent, since the OPE of the picture changing operator with itself
has a double pole.
A resolution to these problems was suggested by using a picture zero
NS string field and an overall mid-point insertion (in the NS sector) of the
double inverse picture changing
operator $Y_{-2}$~\cite{Preitschopf:1989fc,Arefeva:1989cm,Arefeva:1989cp}.
In the most common formulation this operator is built using the doubling
trick as a bi-local operator, composed of two insertions of the inverse
picture changing operator $Y$,
\begin{equation}
\label{Y-2}
Y_{-2}=Y(i)Y(-i)\,,\qquad Y(z)=c\partial \xi e^{-2\phi}(z)\,.
\end{equation}

While this theory does not suffer (in the NS sector) from collisions of
picture changing operators in gauge transformations and in the evaluation of
tree amplitudes, it was criticized for the use of the picture changing
operators on the ground that they support non-trivial kernels. The problem
with the kernels is that they (naively) imply that the equation of motion
derived from the action is not equivalent to the desired one\footnote{Another
formulation of superstring field theory was given by
Berkovits~\cite{Berkovits:1995ab}. This formalism
avoids picture-changing operators altogether and for this reason is
often considered as a more reliable one. Nonetheless, it was recently
found that, up to some subtleties, the two formalisms are classically
equivalent~\cite{Fuchs:2008zx,Kroyter:2009zi}.},
\begin{equation}
Y_{-2}(QA+A^2)=0 \quad \stackrel{?}{\Longleftrightarrow} \quad
QA+A^2=0\,.
\end{equation}

The states in the kernel of $Y_{-2}$ can be characterized by specific
mid-point insertions. It is not clear if such states should be allowed as
part of the space of string fields.
On the other hand, the opposite is also not obvious.
The lack of understanding of the space of string fields is one of the biggest
open problems in the realm of string field theoretical research. It is also
one of the most ignored ones.
While we do not know how to solve this problem, there are several
observations that can be made regarding this space. We discuss and speculate
on this issue in section~\ref{sec:stringFields}.
The following are some of the conclusions we draw,
\begin{itemize}
\item The space of string fields is part of a star algebra.
\item Finite superpositions of regular wedge states with local insertions
      in their interior are part of the space of string fields.
\item Insertions on the boundary and especially at the mid-point are
      problematic and should be discarded in many cases, if not always.
\item Mid-point insertions over string fields, if allowed at all, can only
      be of operators of zero conformal weight.
\item Mid-point insertions in the definition of an action play a different
      (in some sense complementary) role than mid-point insertions over
      string fields.
\item The space of string fields depend on the theory at hand.
\end{itemize}

From the discussion of~\ref{sec:stringFields}, it seems reasonable that
in the cubic theory the problematic mid-point insertions should not be
allowed. Hence, we believe that the cubic theory can be trusted in the NS
sector. Nonetheless, one cannot discard mid-point insertions when the Ramond
sector is included, since the gauge transformations then include
explicit mid-point insertions~\cite{Preitschopf:1989fc}.
These insertions do not belong to the kernel of $Y_{-2}$, but they lead to
divergences that render the theory inconsistent~\cite{Kroyter:2009zi}.
One may still consider the
theory, claiming that it has no gauge symmetry, but we believe that this
theory would not be consistent at the quantum level and at any rate it cannot
be the field theory of open superstrings.

Recently, Berkovits and Siegel suggested to extend by a non-minimal
sector the space of operators on which the string field can
depend~\cite{Berkovits:2009gi}.
From a first-quantized point of view this sector is trivial, since it
decouples by the quartet mechanism~\cite{Henneaux:1992ig}.
It was then suggested to construct a string field theory which includes
this non-minimal sector, where the $Y_{-2}$ insertion is replaced by
a measure factor $\mathcal{N}_\rho$, with $\rho$ being an arbitrary
parameter.
This factor has no kernel and its functional integral over the non-minimal
sector reduces to $Y_{-2}$, regardless of the value of $\rho$.
We give more details of this construction in section~\ref{sec:NMSSFT}, where
we see that the non-zero conformal weight of $\mathcal{N}_\rho$ poses some
difficulties. We further demonstrate that despite that the theory might still
make sense (in the NS sector).

The motivation of the construction~\cite{Berkovits:2009gi} was the problem
with the kernel of $Y_{-2}$. As we argue in section~\ref{sec:stringFields},
this is not really a problem. In any case, if one adopts another point of
view regarding the space of string fields, one should be aware of the fact
that in addition to the potential problems with the kernel of a mid-point
insertion there are potential problems with the space of operators whose OPE
with the insertion is singular. Hence, a formulation with a regular mid-point
insertion, such as~\cite{Berkovits:2009gi,Berkovits:2005bt} is not different
conceptually from the one with the $Y_{-2}$ insertion.

In section~\ref{sec:classical}, we further check the non-minimal theory by
studying the counterparts of known analytical solutions of the minimal
theory. All the solutions
of the minimal theory are automatically solutions of the non-minimal one.
We illustrate that the ``tachyonless tachyon solution'' has the same
properties in the non-minimal theory as in the minimal one.

In section~\ref{sec:lessMin}, we return to the problem of a non-zero-weight
insertion and offer for it another, more trustworthy resolution.
This resolution is based on further extending the space of conformal fields
on which the string field depends by another quartet. We choose this quartet
in a way that avoids the introduction of new zero modes. Then, we modify the
mid-point insertion, including in it dependence on the new quartet variables.
The new mid-point insertion still reduces to $Y_{-2}$ upon integration over
the complete set of non-minimal variables, while having a zero conformal
weight.

The extension of the non-minimal theories to the NS$-$ and to the Ramond
sectors is studied in section~\ref{sec:Ramond}.
We find out that while there is no problem with defining the NS$-$ sector,
the extension of the formalism to the Ramond sectors is not trivial and is
bound in any case to suffer from singularities in its
finite gauge transformations, analogous to those of the minimal theory.
We conclude with some closing remarks in section~\ref{sec:conc}.

\newpage
%===================================
\section{The space of string fields}
%===================================
\label{sec:stringFields}

Here, we present some thoughts regarding the form of the space of string
fields\footnote{There are several types of ``string fields'', e.g., the
``physical'' string fields that appear in the action and the gauge string
fields. Here, we refer to the physical string fields simply as string
fields.}. These are preliminary ideas that try to sum up our knowledge
of the form of known solutions and other legitimate string fields, as well as
the basic algebraic structures and the restrictions that they pose.
While this problem is largely ignored, it is nevertheless important and
is relevant also in the context of the recent development that followed
Schnabl's solution~\cite{Schnabl:2005gv}. In particular, the requirement
that the equations of motion are obeyed even when contracted with the
solution itself and the role of the ``phantom piece'' at the limit
$n\rightarrow \infty$ in this respect~\cite{Okawa:2006vm,Fuchs:2006hw} are
important for proving Sen's first conjecture~\cite{Sen:1999mh,Sen:1999xm}.
Also, the evaluation of the cohomology around the solution specifically
assumed particular behaviour of the allowed string
fields~\cite{Ellwood:2006ba}.

The exact definition of the space of string fields is one of the biggest
unknowns in string field theory. While it is common to use terms such as
``Hilbert space'' in this context, it is not clear if the desired space is a
Hilbert space, since the natural inner product of string fields is not
positive definite, due to the existence of the ghosts\footnote{One might
choose, following~\cite{Ellwood:2001py}, an arbitrary ``reasonable'' norm for
defining the space of string fields and hope that the (topological)
space it defines does not depend on the particular choice of this norm.
Topology is indeed the required structure,
since this is all that is needed in order to define convergence and limits.
Nonetheless, we do not believe that this is a sensible way to define it,
since it is not clear to us that any ``natural norm'' is sensible. See also
the discussion below on separating the matter and ghost sectors.}.
One may think that it is possible to use the basis elements of the
``Hilbert space'' without any constraints on the coefficients.
This is not the case.
To see that consider the insertion
\begin{equation}
\label{BadInsertion}
\cO(z)\ket{0}=\sum_{n=0}^\infty \frac{z^n}{n!}\cO^{(n)}(0)\ket{0}.
\end{equation}
It there were no restrictions on the coefficients, such a state with $|z|>1$
(in the upper half plane coordinates) could have been produced.
This would result in an insertion of the operator $\cO$ not in the interior,
but at the local coordinate patch reserved for the test string field with
which this one is to be contracted.
The test string field can be an arbitrary element of the space of string
fields and in particular it might have an operator insertion at this point,
whose OPE with $\cO$ is singular. Hence, the insertion at $|z|>1$ is bound to
produce singularities in the star product.
The star product of any two elements of the space of string fields should
be well defined in order for the action to make sense. Hence, states
of the form~(\ref{BadInsertion}) with $|z|>1$ should be discarded.
We conclude that the linear combinations should be somehow constrained,
despite the absence of a natural (positive definite) norm on the
``Hilbert space''.

Local insertions at $|z|<1$ should probably be allowed. What about the case
$|z|=1$? Multiplying a string field containing a local insertion $\cO_1$ at
$e^{i\theta}$ by another string field, with an insertion $\cO_2$ at
$e^{i(\pi-\theta)}$ results in the collision of the operators $\cO_i$.
This product is singular, unless the two operators have a regular OPE.
A possible resolution of this specific issue might be to ``divide the
boundary between the state we consider and the test state'', allowing
operator insertions only at $\theta<\frac{\pi}{2}$ or at
$\theta<\frac{\pi}{2}$. This prescription is asymmetric and is hence,
inconsistent with the reality condition of the string field. The reality
condition implies that if an insertion is allowed at $e^{i\theta}$, it
should also be allowed at $e^{i(\pi-\theta)}$. An immediate corollary is
that a minimal requirement from a local insertion on the boundary is to
have a regular OPE with itself. One can still speculate that the range
$0<\theta<\frac{\pi}{2}$ can be divided to segments, with different
insertions, whose mutual OPE's are singular, allowed at different segments.
While this construction naively allows for the coexistence of mutually
inconsistent insertions, it does not respect the
global symmetries generated by\footnote{These ``residual Virasoro
symmetries'' were introduced by Witten in~\cite{Witten:1986qs}.
It was shown by Hata (and reported in the appendix of~\cite{Feng:2001uy}),
that on solutions these global symmetries reduce to gauge symmetries.}
\begin{equation}
K_n\equiv L_n-(-1)^n L_{-n}\,.
\end{equation}
Hence, it seems more reasonable to disregard these unnatural splittings.
It can be seen from this discussion that the mid-point should be treated
differently than the other points with $|z=1|$, since it is invariant
under the star product and is not related to the other points by the
global $K_n$ symmetries. We return to this issue below.

A very natural expectation is that the space of string fields is contained in
a star-algebra. A possible objection to this point of view may rely on the
fact that (for the cubic string field theories such
as~\cite{Witten:1986cc}\footnote{Indeed, our line of argument is even
stronger for the case of the non-polynomial theory~\cite{Berkovits:1995ab},
in which the string field carries zero ghost (and picture) numbers.})
the string field is classically
restricted to be of ghost number one. The set of ghost number one string
fields is not closed under the star product. Moreover, in the action and in
the equation of motion only one or two star products appear. Hence, the
requirement of forming a star algebra is not justified. 
We oppose this line of argument for several reasons:
\begin{itemize}
\item While classically the string field is indeed of ghost number
one, quantum mechanically its ghost number is not
restricted~\cite{Bochicchio:1986zj,Bochicchio:1986bd,Thorn:1986qj}.
\item The space of gauge string fields is (classically) of ghost number zero.
Gauge transformations can be iterated. Hence, the space of (finite) gauge
string fields is certainly closed.
\item The star product is the primary algebraic entity of all the
(open) string field theories, regardless of their polynomiality properties
or ghost number restrictions. It seems natural to exploit the algebraic
structure that is formed by the star product for defining the space of
string fields\footnote{One can consider the even more restrictive
possibility, that the space of string fields forms a $C^*-$algebra.
If this is indeed the case, it might be possible to exploit this
structure for gaining a better understanding of the space of string
fields~\cite{LeonardoTalk}.}.
\end{itemize}
Note, however, that even in the quantum theory, where the ghost number is
not restricted, the space of string fields differs from the algebra of gauge
string fields. While the later is obviously a star-algebra, the former is
obviously not one. The reason is that the space of string fields is odd.
Hence, a product of two string fields is not itself a string field, since
it is even. The gauge string fields can form a star algebra, since they are
always even. The parity is always the same in both cases, regardless of the
ghost number. This is achieved by working with odd coefficient fields.
The coefficient fields are odd for a string field at an even ghost number
and for a gauge string field at an odd ghost number.
We suggest that the algebra one should consider consists of string fields
at an arbitrary ghost number and whose parity is arbitrary. The even
sub-algebra of this algebra forms the (quantum) space of gauge string fields,
while the odd part of this space forms the sought after space of string
fields. Both these spaces can be further restricted by imposing
reality conditions.

The other algebraic structures that we have other than the star
product are the integral and the derivation $Q$ (BRST charge).
We should demand that our space is invariant under the action of $Q$ and
that integrating any of its elements gives a finite result
(i.e., we do not want to consider string fields with diverging action).
Further, we assume that an identity element exists within this algebra.
While it is clear that the identity string
field exists as a functional over the space of string fields, as it
implements the one-vertex, it is also known that this string field is of a
somewhat singular nature. This stems exactly from its role as
the one-vertex: It is a state of zero width, constructed for implementing
a delta-functional gluing. Because of this nature, attempts to construct
solutions based on the identity string
field~\cite{Takahashi:2001pp,Kluson:2002ex,Takahashi:2002ez,Drukker:2003hh}
were not successful in reproducing a calculable action,
although some recent results in favour of these solutions were given
in~\cite{Kishimoto:2009nd}.
Insertions on the identity are always problematic, since in its
geometric picture, the identity string field has no width.
Hence, an insertion on the identity is necessarily an insertion on
the boundary, which should always be handled with care.
Moreover, the fact that the identity string field does not have
a local coordinate patch in its interior (since it has no interior),
results in some paradoxes, when insertions on it are
considered~\cite{Rastelli:2000iu}.

An illustration for the problematic role of insertions on the identity
can be seen as following.
By allowing insertions on the identity one can write the quadratic term
around a solution as a commutator. Moreover, one can eliminate the quadratic
term altogether~\cite{Horowitz:1986dt}.
To that end we introduce (in the cylinder coordinates) the string
field\footnote{The identity based
solutions are generalizations of this construction.},
\begin{equation}
\label{JstringField}
J=\int_{i\infty}^{-i\infty} \frac{dz}{2\pi i} J_B(z) \ket{\id},
\end{equation}
obeying
\begin{equation}
J^2=0\,.
\end{equation}
By definition we can write
\begin{equation}
\label{QasJ}
Q A=[J,A]\,.
\end{equation}
It is clear that the kinetic term around a solution can now be written
as
\begin{equation}
\label{QJeq}
\Q \tilde A=Q \tilde A+[A,\tilde A]=[A+J,\tilde A]\,.
\end{equation}
The action can be written as
\begin{equation}
\label{QJaction}
S=-\int \Big(\frac{1}{2} A[J,A]+\frac{1}{3}A^3\Big )=
   -\frac{1}{3} \int (A+J)^3\,.
\end{equation}
From~(\ref{QJeq}) one can read that the solution
\begin{equation}
A=-J\,,
\end{equation}
corresponds to a zero kinetic term and hence also to a trivial cohomology.
On the other hand,~(\ref{QJaction}) implies that the action of this solution
is zero, in contradiction with Sen's conjecture.
Hence, while it may be useful at times to use~(\ref{QasJ}) for performing
calculations, $J$ of~(\ref{JstringField}) should not be considered a genuine
string field.

Regardless of the issue of
legitimacy of states formed by insertions on the identity string field,
we may add a formal identity to the algebra, which is needed for
defining the finite gauge transformations. It seems to us, that when
one prohibits insertions on the identity string field, there is no
problem with identifying the formal identity with the identity string
field. Note, however, that inconsistencies may arise with derivations of
the star product that do not annihilate the identity. Such derivations should
be discarded when the identity string field is considered. Luckily, the
indispensable derivation $Q$, as well as $\eta_0$ in the supersymmetric
case, annihilate the identity.

To summarize the discussion so far, we are looking for a space that forms a
star-algebra with an identity, i.e., within this space all the relevant
axioms should be obeys, e.g., there are no associativity
anomalies~\cite{Horowitz:1987yz,Bars:2002bt}, $Q$ acts on elements of this
space as a derivation, etc...

One may at this stage wonder whether a set obeying all the above exists at
all. The answer to this question is affirmative. A trivial example would be
the one-dimensional linear space spanned by the identity element.
This is obviously not enough, since we should allow the unintegrated vertex
operators acting on the vacuum in our algebra. The minimal algebra containing
these string fields is the finite span of integer wedge
state~\cite{Rastelli:2000iu,Furuuchi:2001df,Rastelli:2001vb,Schnabl:2002gg}
($W_n$ for $n\in \mathbb N_0$, with $W_0=\ket{\id}$ and $W_1=\ket{0}$)
with insertions of unintegrated
vertex operators at the integer points within the wedge state.
An obvious generalization is the space $\cA$ spanned by a finite linear
combinations of arbitrary finite wedge states
($W_n$ for $0\leq n <\infty$),
with a finite number of local insertions in the interior.
The space $\cA$ seems like a step towards the correct direction, but it also
cannot be the final answer, since it neither contains the
tachyon vacuum~\cite{Schnabl:2005gv}, nor is it closed under finite gauge
transformations.

It is natural to assume that the algebra of string fields is
maximal, i.e., it cannot be further extended.
This requirement might also be important for being able to apply a variant
of the fundamental lemma of the calculus of variations, without which it is
impossible to derive the equations of motion from the action.
A maximal space extending $\cA$ exists. To prove that consider the set of
all (legitimate\footnote{One should define more accurately this
``set of all possible extensions'', or it would neither be a set, nor would
it be well defined. It is clear that we should only have as basis elements
the ones that we already introduced. What is still needed is a notion of
topology, in order to define convergence and equivalence classes of infinite
sums of these elements. The ordering one should consider then should take
into account both the elements of the space, as we suggest here, as well as
the topology of the space. The ordering of topologies can again be achieved
by set inclusion, so Zorn's lemma can still be used.
Here, we ignore this issue for simplicity.}) extensions of $\cA$. This is a
partially ordered set, with set inclusion defining the ordering, i.e.,
\begin{equation}
\cAA_1 \leq \cAA_2 \quad \Longleftrightarrow \quad \cAA_1 \subseteq \cAA_2\,.
\end{equation}
The fact that the partial ordering is given by set inclusion implies that
every chain has an upper bound. Hence, by Zorn's lemma a maximal extension
exists, as stated.

While we can use Zorn's lemma in order to prove that a maximal
extension exists, this form of a proof is as far away from a constructive
argument as it gets. What we would really like to have is a clear criterion
to apply to string fields, in order to determine whether they are legitimate
ones or not.
Moreover, Zorn's lemma proves the existence of the
extension, but not the uniqueness thereof. In fact, the maximal extension is
not unique and further physical data should be used in order to decide among
the possible maximal extensions. In order to demonstrate the non-uniqueness,
let us return to the role of
mid-point insertions. Such insertions, if allowed at all, should only be of
operators of zero conformal weight.
This fact can be understood in the unit disk coordinates.
In this coordinates the conformal transformation
used in the definition of the star product is
\begin{equation}
\zeta \rightarrow \zeta^{2/3}\,,
\end{equation}
while the string mid-point is at $\zeta=0$.
The conformal transformation of the mid-point gives a proportionality factor
of
\begin{equation}
\left(\frac{2}{3}\ep^{-1/3}\right)^h \xrightarrow[\ep\rightarrow 0]{}\left\{
\begin{array}{ll}
\infty & \qquad h>0 \\
0 & \qquad h<0 \\
1 & \qquad h=0
\end{array}
\right .\,.
\end{equation}
The case $h>0$ should be excluded, since it would result in a divergent
star product with all the Fock space states, which are certainly legitimate
ones.
For negative conformal weight, the star product of such string fields
with other string fields is zero.
While this is not a severe problem as the other case, it seems that
the inclusion of such states will invalidate the derivation of the
equations of motion from the action, since no generalization of the
fundamental lemma of calculus of variations can exist for such
states\footnote{A non-zero and non-singular result could presumably be
obtained from the product of two string fields with
$h<0$ mid-point insertions, provided that some of the OPE's of the
insertions with themselves have exactly the needed singularities for
balancing the zeros of the conformal transformations.
One then would have to verify that the resulting string fields still
have regular OPE's with the themselves and with the former ones and so on.
One could then allow arbitrary regular insertions in the interior.
We still feel that such string fields should be
ruled out, since then a particular insertion would be allowed only on a
particular wedge, which seems unnatural.}.

Examples of operators with $h=0$, which can a-priori be inserted at
$z=i$ on string fields in $\cA$, include among others the
operators\footnote{Note that the picture changing operators, whose conformal
weight is also zero cannot be included, since they have singular OPE's with
themselves. Since the mid-point is invariant under the star product, star
multiplying two states with the same picture changing operator insertion at
the mid-point (say, squaring a given string field containing an $X$
insertion) will result in singularities of the star algebra, which we do not
allow.},
\begin{equation}
\label{O1O2example}
\cO_1=\xi\,,\qquad \cO_2=\eta c\,.
\end{equation}
Each of these operators has regular (in fact zero) OPE's with itself.
Thus, each one of them can be included
in an extension of $\cA$ ($\cA \oplus \xi \cA$ for example is a legitimate,
non-maximal algebra). Nevertheless, we cannot include both, since the
OPE $\cO_1\cO_2$ is singular. This proves our assertion that the maximal
extension of the algebra is not unique.
Similar considerations regarding OPE's apply also for operator insertions
at other points of the boundary. On the other hand, there are no
restrictions on the conformal weight for non-mid-point boundary operator
insertions.

More restrictions on the type of allowed string fields can come from
specific properties of a given string field theory.
The non-polynomial superstring field~\cite{Berkovits:1995ab} theory contains
in addition to $Q$ also $\eta_0$ as a derivation. The algebra should be
closed and non-anomalous (i.e., obey Leibniz' law) with respect to $\eta_0$
as well. The modified cubic superstring field
theory~\cite{Preitschopf:1989fc,Arefeva:1989cp,Arefeva:1989cm} contains a
mid-point insertion
of the picture changing operator $Y$. Hence, for this theory the operator
$\cO_1$ of~(\ref{O1O2example}), whose OPE with $Y$ vanishes, may be
considered as a legitimate insertion, while the operator $\cO_2$, whose OPE
with $Y$ diverges, may not\footnote{Note that the fact that $\cO_1=\xi$
does not belong to the small Hilbert
space does not imply that it cannot be used here.
It only implies that it cannot
be inserted arbitrarily. A string field of the form
$(\xi(i)-\xi_0)\ket{V}$ that belongs to the small Hilbert space is an
example of an allowed element in an extension of $\cA$.}.
Note the differences between mid-point insertions on a string field and a
mid-point insertion in the action, such as the $Y_{-2}$ in the cubic
theory:
\begin{itemize}
\item While the former has to have regular OPE with itself, there is no
      such restriction for the action-insertion, since it cannot be
      iterated.
\item While the former has to have $h=0$, the later might have a non-zero
      $h$, provided that the theory is thought of as being some sort of a
      limit. Again, this stems from the fact that there is only one
      such insertion. Thus, a formal zero or infinite constant multiplying it
      can fix the problems with the conformal weight for all string fields.
      This is not the most elegant construction, but it should
      not be ruled out automatically. This situation was encountered in the
      construction of vacuum string field
      theory~\cite{Rastelli:2001uv,Hata:2001sq,Gaiotto:2001ji}. In
      section~\ref{sec:NMSSFT}, we deal exactly with this case in the
      study of the Berkovits-Siegel proposal~\cite{Berkovits:2009gi}.
\item Allowing for a mid-point insertion in the action prohibits mid-point
      insertions in the space of string fields of operators whose OPE with
      the action-insertion is singular or zero. The problem with the
      operators in the kernel of the mid-point action-insertion is known
      and it was the main ground for criticizing the cubic
      theory of~\cite{Preitschopf:1989fc,Arefeva:1989cp,Arefeva:1989cm}.
      We do not think that this is a problem. Rather, as stated,
      it tells us that such states cannot be considered as part of the
      string field algebra, since, again, the fundamental lemma of
      calculus of variations would not hold for them. The problem with
      the operators whose OPE with the mid-point action-insertion is
      singular was mostly overlooked in the literature. Nevertheless,
      string fields with these insertions cannot be allowed, since their
      action is not well defined. In particular it means that in the space
      of string fields mid-point insertions should be restricted not only in
      theories with mid-point insertion with a local kernel, such
      as~\cite{Preitschopf:1989fc,Arefeva:1989cp,Arefeva:1989cm}, but also
      in theories whose mid-point insertion has no kernel such
      as~\cite{Berkovits:2009gi,Berkovits:2005bt}.
      One might now be worried about all theories with mid-point insertions,
      since it seems that different choices
      of the action-insertion give different restrictions on the space of
      string fields. We do not believe that this is a problem, since:
      \begin{itemize}
      \item As we already claimed, mid-point insertions, if allowed at all,
            should any way be constrained, and there is more than one way to
            do that.
      \item The mid-point insertions might be thought of as gauge equivalent
            to more benign ones, but being at the verge of the gauge orbit.
            If this interpretation holds, it is not important if the
            mid-point insertion itself really exists, or is just a singular
            limit, since the gauge orbit exists anyway.
      \item It is perfectly sensible to have the space of string
            fields depend on the particular theory one studies. We turn now
            to a somewhat different example that illustrates this point.
      \end{itemize}
\end{itemize}

Consider a superstring field theory with an R+ sector. The NS sector should
contain states which look like the star product of two R+ states, which in
the most simple case take the form of R+ insertions on the (NS) vacuum. The
NS state then looks like a bi-local R+ insertion on $W_2$. One might
deduce that such states are inherent in any description of the NS+ sector.
Now, we may repeat the above, only with an R$-$ sector. It seems
improbable that both NS+ states based on R+ and NS+ states based on
R$-$ coexist, since they are not mutually local. This problem does not arise
in the worldsheet description, since there the two insertions are inserted at
the same place and one is left, after evaluating the OPE, with a genuine NS+
state. In string field theory on the other hand the insertions are at
distinct points and it might well happen that after multiplying two NS+
states one is left with two R$\pm$ states, which are closer to each other
then to any other insertion.
There might be several resolutions to this difficulty. We believe that the
correct one is to assume that the exact content of the NS+ sector depends on
the theory at hand. The space of string fields is different for the NS+
sector when it is coupled to R+, to R$-$, or to NS$-$.

Ramond sector insertions are solitons from the perspective of the NS sector.
It is a particular case of an allowed infinite linear combination of basis
elements.
We already gave another such example, that of finite gauge transformations.
Yet another explicit case of infinite linear combinations arises when we
consider the wedge states, which as stated already have to be part of the
algebra.
Wedge states are particular cases of surface
states~\cite{LeClair:1989sp,LeClair:1989sj,Rastelli:2001vb}. Surface states
can be defined by requiring that the integral of their star product with any
test state is given by the CFT expectation value of this test state on
a particular surface\footnote{This might sound as a bad definition,
since the expectation value in a boundary CFT does not depend on the specific
disk on which it is evaluated, as they are all conformally equivalent.
However, there are two conformal
maps here. One that defines the surface and the other that defines the
way the test state is inserted (or rather defines its local coordinate
patch). One of these conformal maps can be set to be an arbitrary function,
while the other defines the surface state.}.
Hence, a surface state is characterized by a deformation of the canonical
disk. Infinitesimal deformations are described by the action of
negative Virasoro generators on the vacuum. Thus, a surface state, which is
characterized by a finite deformation, has to have the form,
\begin{equation}
\label{wedge}
\ket{S}=e^{\sum_{n=2}^\infty v_n L_{-n}}\ket{0},
\end{equation}
where the $v_n$ are the coefficients defining the surface
state\footnote{For example, it was shown in~\cite{Rastelli:2000iu}, that
the wedge states $W_n$ have only even non-zero
coefficients and the first non-vanishing one is ($W_n$ here is $\ket{n+1}$
of~~\cite{Rastelli:2000iu}), $v_2=-\frac{(n-1)(n+3)}{3(n+1)^2}$. Note that,
like all the other coefficients, it vanishes for $n=1$, since $W_1$ is the
vacuum state.}.
Expending the exponent in~(\ref{wedge}), we see that any surface state can
be written as an infinite sum of local insertions. Thus, our $\cA$ is
already a space that can be characterized by infinite linear combinations.

Extending $\cA$ so as to include also other surface states is a very natural
direction. The simplest surface state algebra extending that of the wedge
states is the star algebra formed from (hybrid) butterfly states and wedge
states~\cite{Gaiotto:2001ji,Schnabl:2002ff,Gaiotto:2002kf,Fuchs:2002zz,
Uhlmann:2004mv,Fuchs:2004xj}. The states of this algebra can be
characterized by (generalized) Schwarz-Christoffel maps with angles, which
are multiples of $\frac{\pi}{2}$, in the cylinder frame. Other algebras,
naturally extending the two previous ones, can be defined using other basic
angles. The space of finite sums of elements of these spaces (presumably
with a finite number of local insertions) forms the algebra of (generalized)
polygons~\cite{Fuchs:2004xj}. This is still a ``small'' algebra as compared
to the full algebra of surface states (with insertions).

Considering~(\ref{wedge}) for defining surface states, one should note
that the deformation might not only create a new surface, but also reduce an
existing part of the surface. The simplest example for that is the identity
string field, in which all the surface other than the local coordinate patch,
reserved for the test state, was eliminated. Going further to wedge states
with $n<0$ is possible as far as calculating the coefficients is concerned
and there is no obvious problem with the form of these coefficients as a
function of $n$. Nonetheless, it should be clear from the discussion above
that these string fields should be rules out.
One might still hope to be able to see some clue for a singularity when
expanding the coefficients.

Let us illustrate that it is impossible to decide whether a
string field is legitimate or not by the behaviour of the
coefficient of various fields as a function of the level.
Consider the generalization of wedge states to
``unbalanced wedge states''~\cite{Schnabl:2002gg}. These are string fields
which look like wedge states when restricted to the matter or to the ghost
component. However, the wedge states are different in the two sectors.
Writing the states is straightforward, due to the factorization,
\begin{equation}
L_n=L_n^{\mat}+L_n^{\gh}\,,
\end{equation}
and to the fact that the Virasoro operators of different sectors commute,
\begin{equation}
[L_n^{\mat},L_m^{\gh}]=0\,,\qquad \forall n,m\,.
\end{equation}
It was shown that unbalanced wedge states lead to inconsistencies and should
probably be discarded from the space of string fields~\cite{Schnabl:2002gg}.
The coefficients of an unbalanced wedge states cannot behave worse than
those of the two genuine wedge states defining it. We see that behaviour
as a function of the level cannot guide us, as stated.
One might now think that it is possible to solve the problem simply by
discarding unbalanced surface states from the algebra. This is
probably too restrictive, since there are other
unbalanced surface state that do not suffer from the problems of the
unbalanced wedge states~\cite{Fuchs:2005ej}.
Also, with the insertions on the surface states it might be non-trivial to
distinguish linear combinations of balanced and non-balanced surface states.
There is, however, another
important lesson that we should study from this example:
In string field theory there is never a genuine matter-ghost factorization.

%=============================================
\section{Non-minimal superstring field theory}
%=============================================
\label{sec:NMSSFT}

Here, we recall the recent proposal of a non-minimal superstring field
theory~\cite{Berkovits:2009gi}. This formulation uses a mid-point insertion
with a non-zero conformal weight. We describe this
theory and suggest that the problems related to the non-zero conformal
weight of the insertion might be resolved by defining the theory using a
limiting procedure.

The standard cubic open superstring field theory is defined in the NS sector
by the action~\cite{Preitschopf:1989fc,Arefeva:1989cp,Arefeva:1989cm},
\begin{equation}
S= -\int Y_{-2}\Big(\frac{1}{2}A Q A+\frac{1}{3}A^3\Big)\,,
\end{equation}
where $A$ is an NS+ string field in the zero picture and $Y_{-2}$
is given by~(\ref{Y-2}).
The equation of motion takes the form,
\begin{equation}
\label{EOM}
QA+A^2=0\,,
\end{equation}
where, as explained in section~\ref{sec:stringFields}, we assumed in the
derivation that the string fields behave nicely.

In~\cite{Berkovits:2009gi} it was suggested that the string field
should depend also on the non-minimal set of fields, $(u,v)$ and $(r,s)$.
Of these fields the first pair is a $\beta\gamma$ system with
$\la=-\frac{1}{2}$ and the second pair is a $bc$ system with
$\la=-\frac{1}{2}$.
Hence,
\begin{equation}
\label{h_uvrs}
h(u)=h(r)=-\frac{1}{2}\,,\qquad h(v)=h(s)=\frac{3}{2}\,,
\end{equation}
and the energy momentum tensor of the non-minimal sector is,
\begin{equation}
T_{\NM}=\frac{3}{2}\partial u v+\frac{1}{2} u\partial v+
    \frac{3}{2}\partial r s+\frac{1}{2} r\partial s\,.
\end{equation}
These fields are defined in the upper half plane (in the canonical
representation).
One then also adds the anti-holomorphic fields.
Alternatively, one can use the doubling trick and define the original fields
as holomorphic in the whole plane\footnote{One may
wonder if the $u$ zero modes introduce a notion of picture
similarly to the case of the $\gamma$ zero mode. Of course, one can define
a picture associated with this variable and in some sense we are working
with the zero picture in this sector. This would be particularly
transparent if one would fermionize the $u,v$ system. In any case,
one should not worry about this new picture number, since, as we show next,
the non-minimal sector is redundant, i.e., it does not change the
cohomology.}.

The non-minimal fields decouple from the cohomology of the world sheet
theory by adding to the BRST charge the term
\begin{equation}
Q_0=\oint \frac{dz}{2\pi i}r v\,.
\end{equation}
In order for this term to have the standard ghost number we
choose\footnote{One can make here also other choices for the ghost
charge. We choose this particular option, in order to have later
$gh(\chi)=-1$.},
\begin{equation}
gh(u)=-2\,,\qquad gh(v)=2\,,\qquad gh(r)=-1\,,\qquad gh(s)=1\,.
\end{equation}
The decoupling originates from the fact that using $Q_0$, any closed
field in the non-minimal sector can be written as an exact one.
This is usually referred to as the ``quartet mechanism'', where two
conjugate bosonic modes and two conjugate fermionic modes cancel each
other in the cohomology. Here we have conformal fields, but one can
study their modes separately and the logic holds with minor modifications.

We should also add to the BRST charge a piece of the form $cT_{\NM}$,
so as to ensure the relation
\begin{equation}
\label{TQbNM}
T_{\NM}=[Q_{\NM},b]\,.
\end{equation}
This suggests that we should consider,
\begin{equation}
\label{QnmBad}
Q_{\NM}\stackrel{?}{=}\oint \frac{dz}{2\pi i}(r v+cT_{\NM})\,.
\end{equation}
However, such an ad-hoc definition might spoil the quartet argument and even
the nilpotency property of $Q$. Luckily, there exists a similarity
transformation, generated by
\begin{equation}
\label{R}
R=\oint c\Big(\frac{3}{2} s\partial u+\frac{1}{2}\partial s u\Big)\,,
\end{equation}
that gives a result very close to the above,
\begin{equation}
\label{Qtrans}
Q\equiv Q_{\m}+Q_{\NM}= e^R(Q_{\m}+Q_0) e^{-R}\equiv e^R\tilde Q e^{-R}\,.
\end{equation}
The form of $Q_{NM}$ defined by this transformation is almost identical
to~(\ref{QnmBad}) and is given by,
\begin{equation}
\label{Qnm}
Q_{\NM}=\oint \frac{dz}{2\pi i}\Big(r v+cT_{\NM}+
  \gamma^2\big(\frac{3}{2}s\partial u+\frac{1}{2}\partial s u \big)\Big)\,.
\end{equation}
The existence of the similarity transformation implies that the total BRST
charge is nilpotent and its cohomology coincides with that of the minimal
(RNS) sector.
Note, that the last term in~(\ref{Qnm}) does not contribute to~(\ref{TQbNM}).
Hence, the non-minimal conformal fields are indeed primaries, with conformal
weights that are given by~(\ref{h_uvrs}).

The quartet mechanism proves the equivalence of the worldsheet
theories with and without the non-minimal sector.
At the string field theoretical level one has to append to the $Q$
transformation~(\ref{Qtrans}), also its analogous one for the string field,
\begin{equation}
\label{Atrans}
A \rightarrow e^R A\,.
\end{equation}
The transformation~(\ref{Atrans}) is realized by contour integrals, which
can be deformed. Hence, $QA$ and the powers of $A$ transform in the expected
way.
Other than that we also need the superstring field theory
action~\cite{Berkovits:2009gi}.
The action should be constructed in a way that saturates the zero modes.
The number of zero modes of a conformal field of dimension $h$ is given
by,
\begin{equation}
\#_{ZM}=\left\{
  \begin{array}{cl}
      0 &\qquad h>0 \\
      -2h+1 &\qquad h\leq 0
  \end{array}
 \right.\,.
\end{equation}
The zero modes have to be saturated. Hence, a non-zero expectation value
with the field content we have takes the form,
\begin{equation}
\vev{\delta^2(\gamma)\delta^2(u)c^3 r^2} \neq 0\,.
\end{equation}
The expectation value is non-zero and finite, regardless of the
exact location of the insertions, as long as one keeps similar insertions
away from each other\footnote{It is possible to relax
this condition by replacing, say, $c^3$ by $\frac{c''c'c}{2}$. For the
bosonic insertions we can allow the delta functions to go over each other by
using normal ordering with the ``fermionized'' $\xi,\eta,\phi$ variables
(and similarly for $u$)~\cite{Friedan:1985ge} (see
also~\cite{Belopolsky:1997jz} for a discussion on the geometric meaning
of the delta functions). Here,
following~\cite{Berkovits:2009gi}, we stay with the original variables.}.

In light of the above, one can define a one-parameter family of actions as,
\begin{equation}
\label{SSFTaction}
S_\rho=-\int \cN_\rho \big(\frac{1}{2}AQA+\frac{1}{3}A^3\big)\,,
\end{equation}
where $\rho$ is a parameter.
The invertible measure factor $\cN_\rho$ in defined as a product of
insertions at $\pm i$,
\begin{equation}
\label{measureFactor}
\cN_\rho=\mathscr N_\rho(i) \mathscr N_\rho(-i)\,,\qquad
  \mathscr N_\rho=e^{\rho[Q,\chi]}\,,\qquad
  \chi=uc\,.
\end{equation}
Substituting $Q$ one gets,
\begin{equation}
\mathscr N_\rho=e^{\rho \big(rc+u(\gamma^2+\frac{3}{2}c\partial c)\big)}\,.
\end{equation}

The mid-point insertions of the
non-minimal theory possess a space of operators whose OPE's
with the insertions diverge. Consider as an example the
weight-zero primary field $s\partial s e^{2\phi}$. This field has a
(leading order) pole of order $4n$ in its OPE with $[Q,\chi]^n$.
Allowing a string field carrying this operator as a mid-point
insertion in our theory would lead to an essential singularity in the
evaluation of the action, while without the mid-point insertion in the
action, such string fields could have been considered, since all powers of
this operator are well defined.
Hence, it seems that this theory is not a-priori superior over the theory
with the $Y_{-2}$ insertion.

An important property of the string field $A$ is that it is odd.
At the level of the minimal theory this property is implemented by the GSO
projection. Here, we have a new set of odd fields. If not constrained, these
fields can potentially lead to an even string
field. The simplest possible constraint would have been to impose an even
number of $r,s$ insertions in the definition of the string field.
This would have been a bad choice, since while $Q_0$ exchanges the spaces
containing even and odd number of these insertions, the other part of $Q$
leaves both spaces invariant. A better choice is to modify the GSO projection
such that it would also count the non-minimal fermions in its definition.
Now, all parts of $Q$ exchange the two spaces, the NS+ string field $A$
can be consistently defined and all the axioms are obeyed.

As a first test of the action~(\ref{SSFTaction}), consider the case in which
$\frac{1}{2}AQA+\frac{1}{3}A^3$ is independent on the
non-minimal sector. In this case the measure factor can be integrated out to
give the standard $Y_{-2}$ insertion, regardless of the value of $\rho$.
This results from the fact that, in the non-minimal sector, only the $r$
and $u$ zero modes should be integrated,
\begin{equation}
\begin{aligned}
\label{minimalInNM}
\vev{\cN_\rho}_{NM}=&
 \int dr\!_{-\frac{1}{2}} dr_{\frac{1}{2}} du_{-\frac{1}{2}} du_{\frac{1}{2}}
  e^{\rho \big(rc+u(\gamma^2+\frac{3}{2}c\partial c)\big)}
  e^{\rho \big(\bar r\bar c+
         \bar u(\bar \gamma^2+\frac{3}{2}\bar c\partial \bar c)\big)}\\
  =& \int dr d\bar r \, du d\bar u \,\rho^2 r c \,\bar r\bar c \,
   e^{\rho u \gamma^2} e^{\rho \bar u \bar \gamma^2}
  =c\bar c\, \delta(\gamma^2)\delta(\bar\gamma^2)=Y_{-2}\,.
\end{aligned}
\end{equation}
Here, we first wrote the expectation value in terms of integration over the
(two bosonic and two fermionic) zero modes. Then, we changed the variables
to $r,\bar r$ and $u,\bar u$,
\begin{equation}
r=r_{\frac{1}{2}}+z r_{-\frac{1}{2}}\,,\qquad
\bar r=r_{\frac{1}{2}}+\bar z r_{-\frac{1}{2}}\,,\qquad
u=u_{\frac{1}{2}}+z u_{-\frac{1}{2}}\,,\qquad
\bar u=u_{\frac{1}{2}}+\bar z u_{-\frac{1}{2}}\,,\qquad
\end{equation}
with the Jacobian of the transformation canceling between the bosonic and
fermionic modes and left only the terms that will not vanish after the
$r,\bar r$ integration. Performing the integrals lead to the desired
result. This $\rho$ independent result leads one to believe that the
action with different $\rho$ values are equivalent. Indeed, as $[Q,\chi]$
is exact, $\mathscr N_\rho$ can be everywhere inserted, regardless of the
value of $\rho$ and without formally changing on-shell world sheet
calculations.

While the action of string fields that do not depend on the non-minimal
sector is the same in the minimal and non-minimal theories, solutions
of the former are not necessarily solutions of the later, since the kinetic
operator is different in the two theories. Nonetheless,
when one first uses the similarity transformation~(\ref{Qtrans})
and~(\ref{Atrans}), one gets to a theory with $\tilde Q$ as the kinetic
operator\footnote{Note, that we transform the theory, but assume that the
underlying CFT does not change, e.g., the conformal weights of the various
non-minimal fields remain the same.}.
Other than changing $Q$ and $A$, the transformation also changes
the mid-point insertion, since $[R,Q\chi]\neq 0$. The change in the mid-point
insertion is easily found once we notice that
\begin{equation}
[R,\chi]=0\,.	
\end{equation}
This implies that
the change in $Q\chi$ results only from the transformation of
$Q$~(\ref{Qtrans}). Thus, the only change in $\mathscr N_\rho$ is that the
coefficient of the $u c\partial c$ term is now unity.
This term drops out from the evaluation of the expectation value at the
non-minimal sector~(\ref{minimalInNM}), which gives $Y_{-2}$ as before.
Hence, for this transformed theory
we can deduce the following properties of string fields that do not depend
on the non-minimal fields:
\begin{itemize}
\item Such string fields are solutions if and only if they are solutions of
      the minimal theory.
\item These solutions have the same action in both theories.
\item The cohomology around these solutions is the same in both theories,
      again, due to the quartet argument.
\end{itemize}
The only thing we should still prove in order to demonstrate the classical
equivalence of these two theories is that any solution of the
$\tilde Q$-theory is gauge equivalent to a solution that lives
in the minimal sector.
Let $A$ be a solution in this theory and assume that it is sensible to
decompose it as a sum of insertions over wedge states\footnote{Following
section~\ref{sec:stringFields}, we know that an arbitrary string field
can be written as an insertion over the vacuum state. However, in this
representation the decomposition we propose will not work. Nonetheless, all
known analytical solutions have a natural decomposition of the sort proposed
here. Moreover, we do not really have to insist on a wedge state based
solution. Rather, we have to demand that the solution can be decomposed
to a set of insertions over an indexed set of string fields that can be
treated as independent in some way and whose index is additive under the
star product, such that~(\ref{WedgeExpansion}) holds. This is the case,
for example with the generalizations of Schnabl's solution studied
in~\cite{Rastelli:2006ap,Okawa:2006sn}. Also, the sum that we refer to can
well be an integral, or a superposition of discrete and continuous states.}.
What we mean by that is that the elements $A_n$ are insertions over $W_n$
and the equation of motion of $A=\sum A_n$ decomposes as,
\begin{equation}
\sum_{m\leq n} \tilde QA_m+\sum_{k+m\leq n}A_k A_m=0\,,
 \qquad \forall n\,.
\end{equation}
From the discussion in section~\ref{sec:stringFields}, it seems plausible
to assume that
\begin{equation}
\label{WedgeExpansion}
n_0\equiv\inf\big(\{n|A_n\neq 0\}\big)	>0\,.
\end{equation}
If this is the case then~(\ref{WedgeExpansion}) reduces for $n_0$ to the
linearized equation of motion,
\begin{equation}
\tilde QA_{n_0}=0\,.
\end{equation}
The quartet argument then implies that it is possible to find a (possibly
trivial) linearized gauge transformation that is built over $W_{n_0}$ that
sets the non-minimal part of $A_{n_0}$ to zero.
Consider now~(\ref{WedgeExpansion}) for $2n_0$. While this equation contains
a non-linear part, its non-minimal component is zero, since, when the wedge
state decomposition makes sense and no boundary insertions are allowed,
the star product, as well as $Q$, do not change the number of non-minimal
insertions.
Hence the equation for the non-minimal component of~(\ref{WedgeExpansion})
at level $2n_0$ reads,
\begin{equation}
\sum_{m\leq 2n_0} \tilde QA_m^{(\NM)}=0\,.
\end{equation}
Using the quartet argument again, we can push the non-minimal insertions
to $m\geq 4n_0$ and so on, all the way to infinity. While this construction
is still not a proof that every non-minimal string field is gauge equivalent
to a minimal one, it seems like a strong indication in this direction and
thus, to the classical equivalence of the minimal theory and the non-minimal
one, based on~$\tilde Q$.
Using again the transformation generated by $R$~(\ref{R}), we conclude that
the minimal and non-minimal theories are probably classically equivalent.

In the discussion regarding equivalence we implicitly assumed that the
non-minimal theory is well defined and that the equations of motion can be
derived from its action.
It is not at all obvious that this is indeed the case.
The problems result from the fact that this theory has a mid-point
insertion, which is not of conformal dimension zero.
In fact, $\chi$ and $[Q,\chi]$ are primary fields of dimension
$-\frac{3}{2}$. This renders the theory not well defined.
Specifically, the conformal expectation value, which appears in the
definition of the theory has different meanings on different choices
of the canonical disk. Trying to evaluate the expression on a different
disk will induce a non-trivial transformation on the definition of the
theory, due to the non-vanishing conformal weight. Luckily we can easily
evaluate this change, since not only $[Q,\chi]$ is a (weight $\frac{3}{2}$)
primary, but also $[Q,\chi]^n$ is a (weight $\frac{3}{2}n$)
primary\footnote{This stems from the fact that the fields of the non-minimal
sector form first-order systems.}. This implies that the whole role of the
conformal transformation is to induce a rescaling of $\rho$ by a factor of
$f'^{\frac{3}{2}}$.

The rescaling of $\rho$ might seem to be not that bad. However, it introduces
several problems. First, while it is assumed that the superstring
field theory~(\ref{SSFTaction}) is
independent of $\rho$, it is not a-priori clear that this is indeed the case,
due to the fact that it is an off-shell formalism.
To overcome this problem one should prove the (quantum)
$\rho$-independence.
Another problem is that one should be careful so as to define the
kinetic and potential terms on suitable surfaces, since even if
the theory is $\rho$-independent, it is certainly not the case for each of
the two terms separately. A relative rescaling of the two parts of the action
can be compensated by a rescaling of $A$.
However, here we don't have a rescaling
by a number, but by an operator insertion at the mid-point. As suggested
in section~\ref{sec:stringFields}, one should avoid string fields with
mid-point insertions of non-zero conformal weight operators. Hence, we cannot
compensate the $\rho$ rescaling in the case at hand.
Moreover, the conformal transformations one usually considers in the
context of the star product tend to have a singular derivative at the string
mid-point. Thus, the rescaling we mention actually sends $\rho$ to zero or
infinity. This situation implies that we should treat the theory with care,
regularize it in order to move the insertion away from the mid-point
and at the end take the limit in a way that produces sensible results.

The procedure of regularizing the theory can be criticized on the ground that
the mid-point should probably not be thought of as a limit of the interior
points, as far as, say, defining string fields is concerned. Moreover,
associativity of the star product and gauge invariance are lost
during regularization and are recovered only as the limit is attained.
We ignore these subtleties for now, assuming that our manipulations
can be somehow justified. We offer a cleaner resolution of the problem
in section~\ref{sec:lessMin}.

For the purpose of regularization we implement the mid-point insertion as a
product with the string field $I_{\delta,\ep}(\rho)$, defined in the
cylinder coordinates as the wedge state $W_\delta$ with insertions of
$\mathscr N_\rho$ at $\pm i h\equiv \pm\arctan\big(i(1-\ep)\big)$.
Removing the regularization amounts to sending both $\delta$ and $\ep$
to zero. If we consider string fields which are wedge states $W_{n_i}$ with
finite number of local insertions at $z_i$, associativity is approximately
recovered when $n_i\gg \delta$ and $h\gg \Im(z_i)$.
Up to ``small'' corrections coming from non-associativity, the equations of
motion take the regularized form,
\begin{equation}
\label{BSeomReg}
I_{\delta,\ep}(\rho) (QA+A^2)=0\,.
\end{equation}
If $h\gg \Im(z_i)$ or if there are no insertions on the
boundary\footnote{String fields with ($h=0$) insertions at the mid-point
or line integrals that go all the way to the mid-point (such as the $B$
insertion below) should also be regularized.}, one can safely
take the limit $\delta \rightarrow 0$. One should
not worry about the fact that this limit produces a non-legitimate string
field, since it only appears in a specific way in the equation of
motion and should not be considered as an element of the star algebra.
Now, we can multiply the equation of motion from the left by
$I_{\delta,\ep}(-\rho)$, so as to obtain the correct equation of
motion~(\ref{EOM}). Note that the limit $\ep \rightarrow 0$
($h\rightarrow \infty$) is also necessary, in order to
justify~(\ref{BSeomReg}), since it is only in this limit that associativity,
formally used in deriving this equation, is retained.
Taking the same wedge width during regularization implies the
geometric picture presented in fig.~\ref{fig:Vertices}.
It seems that, at least classically and for this sort of solutions,
the theory can be defined as a limit using this geometrical picture.
\FIGURE{
\label{fig:Vertices}
\epsfig{figure=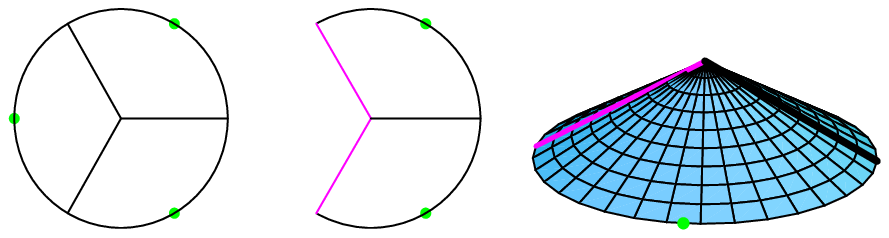, width=15cm}
\caption{The three vertex (left), in its canonical
conformal frame, with the insertions represented by green dots.
The two vertex (middle) should be represented using the same mapping.
The pink lines should be glued.
One cannot use the canonical frame for the two vertex, since it would
induce an infinite rescaling of $\rho$ and hence change the mid-point
insertions.
A more ``intuitive'' picture (right) of the two-vertex is that of a cone
(only one insertion is visible). The black and pink lines separating the
two string fields are shown in this form as well.
One could have decided to use the canonical surface for the two vertex,
in which case the three vertex would have been inserted on a cone with
a $3\pi$ circumference. The values of $\rho$ in the two representations
are formally related by an infinite rescaling.}
}

%============================
\section{Classical solutions}
%============================
\label{sec:classical}

In~\cite{Erler:2007xt}, Erler defined the vacuum solution of the modified
cubic superstring field theory using its formal gauge
representation~\cite{Okawa:2006vm}\footnote{Despite the fact that this
solution lives in the NS+ sector, it describes the vacuum without the
original D-brane.
This led to a proposal of a modified solution with support in the
NS$-$ sector, which might seem more adequate in the context of tachyon
condensation~\cite{Aref'eva:2008ad}. It was claimed in~\cite{Fuchs:2008zx}
that these two solutions are gauge equivalent and the gauge family
that interpolates between the two solutions was constructed. Hence, it
seems natural to assume that Erler's solution describes the non-perturbative
vacuum also on the BPS D-brane, despite the absence of tachyons in this case.
}.
For that, he simply took the same gauge string field as the one defining
Schnabl's tachyon vacuum in the bosonic theory~\cite{Schnabl:2005gv},
\begin{equation}
\label{SchLa}
\La=B c(0) \ket{0},
\end{equation}
where $c$ is a $c$-ghost insertion in the cylinder coordinates
and $B$ is a line integral of the $b$-ghost in the same coordinate system,
\begin{equation}
B=\int_{i\infty}^{-i\infty} \frac{dz}{2\pi i}b(z)\,.
\end{equation}
The solution itself is then given by
\begin{equation}
\label{ErlerSol}
A_{\m}=Q_{\m}\La\frac{1}{1-\La}\,,
\end{equation}
where the subscripts ``$\m$'' on $A_{\m}$ and $Q_{\m}$ remind us that they
are the string field and BRST charge of the supersymmetric theory in the
minimal sector (the same holds for $K_{\m}$ and $T_{\m}$ below).
Substitution gives\footnote{We use the split-string notations
here~\cite{Erler:2006hw,Erler:2006ww},
according to which operators are interpreted as if they are inserted (at the
origin) on the identity string field. Then, all products are star products.
Again, one should not worry about the legitimacy of the string fields that
are built on the identity, which compose the solution, since the solutions
itself is a legitimate string field.},
\begin{equation}
\label{ErlerSolExp}
A_{\m}=A_{\bos}-W_{\frac{1}{2}} B\gamma^2 W_{\frac{1}{2}}
  =W_{\frac{1}{2}} c \frac{K_{\m} B}{1-W_1} c W_{\frac{1}{2}}
    -W_{\frac{1}{2}} B\gamma^2 W_{\frac{1}{2}}\,,
\end{equation}
where $A_{\bos}$ is the bosonic solution and $K_{\m}$ is the line integral
of the energy momentum tensor,
\begin{equation}
\label{Km}
K_{\m}=\int_{i\infty}^{-i\infty} \frac{dz}{2\pi i}T_{\m}(z)\,.
\end{equation}

Actually, strictly speaking the first term in the r.h.s is not the same
as the bosonic solution, the difference being that the later is defined
using $K_{\bos}$ rather than $K_{\m}$. Of course, $K_{\m}$ is not defined
in the bosonic theory, but $K_{\bos}$ is defined in the RNS one (with a
different central charge, due to the different dimensionality). Note,
that the solution depends on $K$ not only via its explicit presence,
but also implicitly in defining the wedge states over which the
solution is constructed,
\begin{equation}
W_n=e^{\frac{n \pi}{2} K} W_0\,.
\end{equation}

We now want to identify the analogous solution in the non-minimal theory.
One way would be to use the equivalence advocated in
section~\ref{sec:NMSSFT}. We prefer to follow another path for the
construction of this solution. Both ways should result in solutions, which
are at least gauge equivalent.

We again define the solution as a formal gauge solution using
$\La$~(\ref{SchLa}), only now we have to
include also $Q_{\NM}$~(\ref{Qnm}) in~(\ref{ErlerSol}). This extra piece has
no effect on the result other than changing $K_{\m}$ to
$K\equiv K_{\m}+K_{\NM}$. Explicitly, the solution is,
\begin{equation}
\label{NMSolExp}
A=W_{\frac{1}{2}} c \frac{K B}{1-W_1} c W_{\frac{1}{2}}
    -W_{\frac{1}{2}} B\gamma^2 W_{\frac{1}{2}}\,,
\end{equation}
Hence, we can say that the solution in this case is exactly
the same as it is in the non-minimal theory, with the understanding that
the geometric interpretation is uniform for all sectors. This is analogous
to the declaration that the first term in~(\ref{ErlerSolExp}) is ``the same''
as the bosonic solution. Note, that despite the discussion in
section~\ref{sec:stringFields} regarding the problems with the unbalanced
wedge states, there would have been no problem of
principle in defining solutions over wedge states, which are
not balanced between the minimal and non-minimal sectors, since the central
charges of both sectors vanish separately.

For evaluating the action of the solution we have to
regularize the insertion as stated. 
Also, in this case one has to regularize the $B$ line integral for the same
reason. However, it seems that this does not pose any problems.
There is no explicit dependence on the non-minimal fields. Rather, they
enter the evaluation of the action only via their presence in 
$\mathscr N_\rho$ and in the geometry of the surface on which the CFT
expectation value is to be evaluated ($K$ is geometrically realized as a
derivative with respect to strip length). This allows us to evaluate the
expectation value in the non-minimal sector first, leaving us exactly the
$Y_{-2}$ insertion. The fact that the same surface state is used in the
minimal and the non-minimal sectors enables us to carry on with this
$Y_{-2}$ with the evaluation in the minimal sector.
Now, we can use the fact that we are left with a zero-weight insertion and
send $\delta$ and $\ep$ to zero. This results in exactly the action of
Erler's solution.

The evaluation of the cohomology around the solution is
even simpler. We only have to note that the same contracting homotopy
operator that works for Schnabl's and Erler's
solutions~\cite{Schnabl:2005gv,Erler:2007xt}, works also here, where again,
the non-minimal sector enters implicitly in the definition of the wedge
states on which this string field is built.
Since the action and the cohomology give the desired results, we conclude
that the non-minimal theory also supports this ``tachyonless tachyon
solution''. This is in accord with our claim that the minimal and
non-minimal theories are classically equivalent.

Similarly to the case of the non-perturbative vacuum one can translate
other solutions to the non-minimal theory. In particular, the analytical
solutions describing marginal
deformations~\cite{Schnabl:2007az,Kiermaier:2007ba,Erler:2007rh,Okawa:2007ri,
Okawa:2007it,Fuchs:2007yy,Fuchs:2007gw,Kiermaier:2007vu,Kiermaier:2007ki}
can be recast in this language.

%=======================================================================
\section{An extended non-minimal formalism with a zero-weight insertion}
%=======================================================================
\label{sec:lessMin}

In section~\ref{sec:NMSSFT}, we discussed the problems with the negative
conformal dimension of $\chi$.
We suggested that if one thinks of the non-minimal superstring field
theory as a limit in some sense of regularized theories, it might be possible
to make sense out of it, since the formally infinite rescaling of $\rho$
between different conformal frames becomes finite when the regularization is
imposed. Nevertheless, it is desirable to find a more elegant resolution of
the problem.
A natural avenue to follow is to further extend the space of operators,
in a way that would enable an $h=0$ insertion.

The most straightforward extension is to declare that $\rho$ itself
is not a parameter, but a conformal field of dimension $h=\frac{3}{2}$.
This cannot be the whole story, since if $\rho$ is to be considered a
genuine conformal field, it should contribute to the energy-momentum tensor
and the total energy-momentum tensor should be reproduced by the relation
\begin{equation}
[Q,b]=T\,.
\end{equation}
The appearance of $\rho$ in $Q$ implies that the cohomology might
have changed. The simplest way to avoid that is to make $\rho$ an
element of a second quartet and define its contribution to $Q$
in a way analogous to~(\ref{Qnm}).
This still cannot be the end of the story, since the field $\sigma$
conjugate to $\rho$ must have $h=-\frac{1}{2}$, which implies that it
carries two zero modes. One could add a third
quartet and eliminate the zero modes of the two new quartets by
including a new piece in $\chi$. However, now
$\chi$ and $[Q,\chi]$ depend both on $\rho$ and on $\sigma$.
Hence, their product at the same point diverges. This can be resolved by
including normal ordering in the definition, but even then, one does
not get the desired result, as can be seen by a direct substitution.

While this modification does not work, there is still a way
out. Let us add to the $(u,v,r,s)$ quartet
a second quartet $(\tilde u,\tilde v,\tilde r,\tilde s)$, with all the
fields having conformal dimension $h=\frac{1}{2}$. The new quartet has
no zero modes.
We now have,
\begin{equation}
Q_0=\oint \frac{dz}{2\pi i}\big(r v+ \tilde r \tilde v\big),
\end{equation}
which we transform using a similarity transformation generated by,
\begin{equation}
\label{Rext}
R=\oint c\Big(\frac{3}{2} s\partial u+\frac{1}{2}\partial s u
  +\frac{1}{2} \tilde s\partial \tilde u
  -\frac{1}{2}\partial \tilde s \tilde u\Big).
\end{equation}
This leads to,
\begin{equation}
Q_{\NM}=\oint \frac{dz}{2\pi i}\Big(r v+\tilde r\tilde v+cT_{\NM}
  +\gamma^2\big(\frac{3}{2}s\partial u+\frac{1}{2}\partial s u
    +\frac{1}{2}\tilde s\partial \tilde u
     -\frac{1}{2}\partial \tilde s \tilde u\big)\Big),
\end{equation}
where the energy momentum tensor is defined as usual for these fields,
\begin{equation}
T_{\NM}=\frac{3}{2}\partial u v+\frac{1}{2} u\partial v+
    \frac{3}{2}\partial r s+\frac{1}{2} r\partial s+
 \frac{1}{2}\partial \tilde u \tilde v- \frac{1}{2}\tilde u\partial \tilde v+
\frac{1}{2}\partial \tilde r \tilde s-\frac{1}{2}\tilde r\partial \tilde s\,.
\end{equation}
Let us now define,
\begin{equation}
\chi=\tilde u^3 u c\,.
\end{equation}
This operator is primary and has zero conformal weight. Moreover, its
$Q$ commutator and all the powers thereof are $h=0$ primaries.
Hence, there are no problems with a mid-point insertion of the operator,
\begin{equation}
\label{LMgoodIns}
\cN_\rho=\cn_\rho(i) \cn_\rho(-i)\,,\qquad
\cn_\rho=e^{\rho[Q,\chi]}=e^{\rho\big(\tilde u^3 (rc+u \gamma^2)
   +3\tilde u^2\tilde r u c\big)}\,.
\end{equation}
We should also define the ghost numbers in a consistent way, e.g.,
demanding $gh(\chi)=-1$. The simplest way is to keep the ghost
numbers introduced so far as before and define,
\begin{equation}
gh(\tilde u)=gh(\tilde v)=0\,,\qquad
gh(\tilde r)=1\,, \qquad gh(\tilde s)=-1\,.
\end{equation}
Integrating the $(u,v,r,s)$ sector leads, as before, exactly to $Y_{-2}$,
regardless of $\rho$'s value. Note, that the last term of $[Q,\chi]$ does
not contribute, since it carries no $r$ zero mode, but does carry a
factor of $c$.
As no dependence on the $(\tilde u,\tilde v,\tilde r,\tilde s)$ sector
remains, it can now be trivially integrated, since it carries no zero modes.

Solutions of the theory based on these two quartets can be defined as before.
For the same reasons as in the one-quartet theory, they will share
the properties of their ``minimal'' counterparts. However, unlike in the
one-quartet case, the $h(\chi) \neq 0$ related difficulties are absent,
there is no need to regularize the theory as part of its definition and
there are no obstacles with more general solutions for which the
regularization can potentially break down.

%======================
\section{Other sectors}
%======================
\label{sec:Ramond}

In this section, we discuss the incorporation of other possible sectors into
the non-minimal theory.
The NS$-$ sector was first incorporated into superstring field theory in
the context of the non-polynomial
theory~\cite{Berkovits:2000zj,Berkovits:2000hf}.
There, it was understood that in order to keep the algebraic axioms of the
star product intact, one should tensor the NS$\pm$ string fields
with the Pauli-matrices
$\sigma_{0,1}$ respectively ($\sigma_0 \equiv {\bf 1}_{2\times 2}$)
and the gauge string fields should be tensored with $\sigma_{2,3}$.
The operators $Q$ and $\eta_0$ are tensored with $\sigma_3$.
The action then should be supplemented
with a factor of $\frac{1}{2}$ and a trace over this ``internal Chan-Paton''
space, in order to obtain the usual result for the NS+ sector. The various
sectors should be put into the string field, presumably tensored also with
genuine Chan-Paton factors, according to the details of the D-brane
system one wants to describe.

The generalization of these ideas to the cubic theory is
simple~\cite{Arefeva:2002mb}. Here, the parity of the string fields
and the gauge string fields is reversed. Hence, the string fields come
with $\sigma_{2,3}$ factors and the gauge string fields get the
$\sigma_{0,1}$. The $Y_{-2}$ insertion in the action comes with a
$\sigma_3$ factor, just like $Q$.
No other changes are necessary.
The case at hand is nothing but a variant of
this theory. Thus, the inclusion of the NS$-$ sector should be
done exactly in the same way, only, as we explained in
section~\ref{sec:NMSSFT}, the GSO projection should be performed over the
minimal and non-minimal sectors together. The sole difference is that the
$Y_{-2}$ insertion is now replaced by the $\cN_\rho$ insertion. Hence, all
will work well provided we assign a $\sigma_3$ factor also to $\cN_\rho$.

In the minimal theory the Ramond sector can (naively) be included by
introducing a picture $-\frac{1}{2}$ Ramond string field and adding to the
action the following term~\cite{Preitschopf:1989fc},
\begin{equation}
\label{RsecMin}
S_R= -\int Y\big(\frac{1}{2}\al Q \al+A \al^2\big).
\end{equation}
Here, $\al$ is the Ramond string field, residing in the R+ or R$-$ sector,
according to the D-brane at hand.
Since the R+, R$-$ and NS$-$ sectors are mutually exclusive,
there can be no need for adding the internal Chan-Paton factors when
the Ramond sector is considered.
One might have thought that the Ramond part of the action should be written
as,
\begin{equation}
\label{RsecNM}
S_R\stackrel{?}{=} -\int \cn_\rho\big(\frac{1}{2}\al Q \al+A \al^2\big),
\end{equation}
where the $\cn_\rho$ factor is inserted at the mid-point.
This cannot be the case, since only one of the zero modes of $u$ is
addressed in this way.
Another possibility could have been to use the same insertion as in the
NS sector, i.e., $\cN_\rho$, while putting on top of it also one picture
changing operator $X$. This, however, would be inconsistent due to
singularities in the OPE of $X$ and $[Q,\chi]$. Redefining the picture of
$\al$ to be $+\frac{1}{2}$ and plugging $Y$ instead of $X$ does not lead to
immediate singularities, but those would nevertheless emerge after the
integration of $u$. Hence, one should look for a less trivial modification
of the insertion in order to define the Ramond sector. We shall not try that,
since even upon establishing a sensible insertion for the Ramond sector,
its gauge transformation would not be well defined.
This would be just the same as for the minimal theory, in which collisions of
picture changing operators occur upon iterating the fermionic part of the
gauge symmetry~\cite{Kroyter:2009zi}. This destroys the finite gauge
symmetry, which we should have if the theory is to describe superstring
theory.
We cannot expect a non-minimal field to replace the $X$ insertion in the
gauge transformation, since a genuine picture changing should be enforced in
order to obtain a string field in the picture that we specified upon
defining the theory. Thus, it seems that the non-minimal theory cannot
resolve the problems with the Ramond sector. Another resolution is needed.
Some preliminary ideas regarding such a resolution were presented
in~\cite{Kroyter:2009zi}.

%====================
\section{Conclusions}
%====================
\label{sec:conc}

This work focused on two complementary issues,
the definition of the space of string fields and the definition of the
(super)string field theory.
While we still do not know how to properly define the space of string fields,
we made some observations regarding its properties.
We hope that these observations would be useful for properly defining this
elusive space.

The observations we made regarding the space of string fields enabled us to
conclude that the common obstruction to the $Y_{-2}$ insertion in the
standard cubic superstring field theory is not well founded.
These observations also enabled us to recognize a problem (non-zero
conformal weight of a mid-point insertion) with the definition of
the alternative formalism~\cite{Berkovits:2009gi}, devised by Berkovits and
Siegel in order to avoid the $Y_{-2}$ insertion.
We managed to show that the non-minimal theory can be modified in a way
that avoids the problems with the conformal weight.

Nonetheless, as we showed in the companion paper~\cite{Kroyter:2009zi},
there is a genuine problem with the minimal formulation. The origin of
this problem is not the kernel of $Y_{-2}$ and its role in defining the
action, but rather the singularity of OPE's of picture changing operators
and their presence in defining finite gauge transformations.
Unfortunately, it seems that the non-minimal theory cannot resolve these
issues.
It would be highly desirable to have a (cubic) theory that can be defined
with string fields with only interior-insertions and whose gauge
transformations do not contain explicit operator insertions. Such a theory
would be well defined in both sectors presumably even at the quantum level.
We are currently studying this possibility.

During our study of the non-minimal theory we verified that the
vacuum solution on a BPS manifold exists also in this theory.
We interpreted that as an evidence in favour of the validity of
the non-minimal formalism. However, one might as well interpret this
observation as standing in favour of the validity of this solution.
Other than by its action and the cohomology around it, a solution can also be
characterized by the boundary state to which it
corresponds~\cite{Kiermaier:2008qu}\footnote{See
also~\cite{Gaiotto:2001ji,Hashimoto:2001sm,Michishita:2004rx,Ellwood:2008jh,
Kawano:2008ry,Kawano:2008jv,Kishimoto:2009cz}.}.
It would be interesting to extend this construction also to the various
superstring field theories. This would further support our
interpretation of the superstring vacuum solution.

\newpage
%=========================
\section*{Acknowledgments}
%=========================

I would like to thank Nathan Berkovits, Ted Erler, Udi Fuchs,
Michael Kiermaier, Leonardo Rastelli and Barton Zwiebach for many
discussions on the issues covered in this work.

It is a pleasure to thank the organizers and participants of the KITP
workshop ``Fundamental Aspects of Superstring Theory'', where a major part
of this work was performed, for providing a very stimulating and enjoyable
environment. While at the KITP, this research was supported by the
National Science Foundation under Grant No. PHY05-51164.
It is likewise a pleasure to thank the Simons Center for Geometry and Physics
and the organizers and participants of the ``Simons Workshop on String Field
Theory'' for a great hospitality and for many discussions on and around the
topics presented in this manuscript.

This work is supported by the U.S. Department of Energy (D.O.E.) under
cooperative research agreement DE-FG0205ER41360.
My research is supported by an Outgoing International Marie Curie
Fellowship of the European Community. The views presented in this work are
those of the author and do not necessarily reflect those of the European
Community.

%=================
\bibliography{bib}
%=================

\end{document}